\documentclass[12pt]{article}
\usepackage[colorlinks=false, pdfborder={0 0 0}]{hyperref}
\usepackage{float}

\usepackage{graphicx,url}

\usepackage[brazil]{babel}   
\usepackage[utf8]{inputenc}  

\usepackage{sbc-template}
\title{Justiça Algorítmica: Instrumentalização, Limites Conceituais e Desafios na Engenharia de Software}

\author{Lucas Rodrigues Valença\inst{1}, Ronnie de Souza Santos\inst{1}}

\address{Department of Electrical \& Software Engineering -- University of Calgary\\
  Calgary -- AB -- Canada
  \email{\{lucas.rodriguesvalen,ronnie.desouzasantos\}@ucalgary.ca}
}

\renewcommand{\refname}{Referências}

\begin{document} 

\maketitle

\begin{abstract}
This article describes ongoing research with the aim of understanding the concept of justice in the field of software engineering, the factors that underlie the creation and instrumentalization of these concepts, and the limitations faced by software engineering when applying them. The expansion of the field of study called ``algorithmic justice''  fundamentally consists in the creation of mechanisms and procedures based on mathematical and formal procedures to conceptualize, evaluate and reduce biases and discrimination caused by algorithms. We conducted a systematic mapping in the context of justice in software engineering, comprising the metrics and definitions of algorithmic justice, as well as the procedures and techniques for fairer decision-making systems. We propose a discussion about the limitations that arise due to the understanding of justice as an attribute of software and the result of decision-making, as well as the influence that the field suffers from the construction of computational thinking, which is constantly developed around abstractions. Finally, we reflect on potential paths that could help us move beyond the limits of algorithmic justice.
\end{abstract}
     
\begin{resumo} 
  Este artigo descreve uma pesquisa em andamento com o objetivo de compreender o conceito de justiça no campo da engenharia de software, os fatores que fundamentam a criação e instrumentalização desses conceitos e as limitações enfrentadas pela engenharia de software ao aplicá-los. A expansão do campo de estudo denominado de ``justiça algorítmica'' consiste fundamentalmente na criação de mecanismos e procedimentos matemáticos e formais para conceituar, avaliar e reduzir vieses e discriminações causadas por algoritmos. Realizamos um mapeamento sistemático no contexto de justiça na engenharia de software, compreendendo as métricas e definições de justiça algorítmica, assim como os procedimentos e técnicas para sistemas de tomada de decisão mais justos. Propomos,,então,o, uma discussão acerca das limitações que surgem devido à compreensão de justiça como um atributo de software e resultado de tomadas de decisões, assim como a influência que o campo sofre decorrente da construção do pensamento computacional, que constantemente é desenvolvido em torno de abstrações. Por fim, refletimos sobre possíveis caminhos que podem nos ajudar a superar os limites da justiça algorítmica.
\end{resumo}

\section{Introdução}
A empresa de jornalismo Propublica publicou, em 2016, uma investigação sobre o \textit{Correctional Offender Management Profiling (COMPAS}), uma ferramenta de avaliação de risco amplamente utilizada nos tribunais dos Estados Unidos, relatando que o sistema rotulava pessoas negras que não possuíam reincidência criminal como alto risco duas vezes mais do que pessoas brancas \cite{propublica2016}. Sistemas de vigilância digital também têm sido alvo de críticas, um caso notável ocorreu no Brasil: os sistemas de reconhecimento facial foram usados como dispositivo de vigilância durante o carnaval, embora não haja evidências de que esses sistemas diminuam significativamente a criminalidade ou que previnam crimes durante o evento brasileiro, resultando na prisão de pessoas inocentes \cite{alencar_2023_racismo_algoritmico}.

Essa crescente consciência de discriminação provocada por algoritmos levou a um crescimento de estudos sobre impactos sociais negativos de máquinas, sistemas digitais e algoritmos, que muitas vezes se estabelecem em conceitos como ``\textit{fairness}", ou justiça \cite{verma2018fairness}. Da mesma forma, a comunidade de pesquisadores da ciência da computação e engenharia de software tem proposto diferentes abordagens para testar quão justo é um sistema \cite{chen2024fairnesstesting, udeshi2018fairnesstesting}. \cite {chen2024fairnesstesting} define ``bugs de justiça'' como ``qualquer imperfeição em um sistema de software que cause uma discrepância entre as condições de justiça existentes e necessárias'' e ``teste de justiça'' como ``qualquer atividade projetada para revelar \textit{bugs} de justiça através da execução de código".

A maior parte desses mecanismos se baseia em procedimentos matemáticos e formais para conceituar, avaliar e reduzir vieses e discriminações causadas por algoritmos. Porém, a maneira como algoritmos integram a sociedade, assim como os objetivos daqueles que os idealizam e desenvolvem e seus impactos, se entrelaçam em dinâmicas sociais que não são estruturadas de maneira estável. A partir dessa compreensão, esta pesquisa se propõe a analisar as limitações das abordagens de justiça algorítmica promovidas no campo da ciência da computação e engenharia de software.

\section{Método}
\subsection{Objetivo da pesquisa}
Este artigo faz parte de uma dissertação de mestrado em andamento. Como primeira etapa da pesquisa, está sendo realizado um estudo terciário sobre o conceito de justiça no campo da engenharia de software, os fatores que fundamentam a criação e a instrumentalização desses conceitos e as limitações enfrentadas pela engenharia de software ao aplicá-los. 

\subsection{Análise}
% Uma revisão de literatura narrativa é uma técnica para construir teorias e gerar hipóteses \cite{baumeister_writing_1997}. 
Neste trabalho, realizamos um mapeamento sistemático terciário em conjunto com uma metassíntese. Metassínteses são definidas como as traduções interpretativas derivadas da integração ou comparação de descobertas entre estudos qualitativos \cite{cavalcante2020, flick_sage_2024}. Como explica \cite{stern1997qualitative}, a metassíntese visa revisar um grupo de estudos ``com o objetivo de descobrir os elementos essenciais e traduzir os resultados em um produto final que transforme os resultados originais em uma nova conceituação”. O estudo é considerado um mapeamento terciário porque mapeia apenas revisões sistemáticas de literatura \cite{kitchenham_systematic_2010}. Por fazer parte de uma pesquisa em andamento, aqui conduzimos apenas um mapeamento sistemático dos 10 primeiros artigos \footnote{Os artigos podem ser encontrados no link a seguir: \href{https://figshare.com/s/01b30a1a7737696b5c2f}{https://figshare.com/s/01b30a1a7737696b5c2f}}, de um total de 60 incluídos na busca realizada, o que nos permite encontrar evidências que direcionem o foco da revisão sistemática que será realizada na dissertação \cite{keele2007guidelines}.

\subsection{Estratégia de busca}
O processo de busca ocorreu de maneira manual e automática. Para a busca automática, uma \textit{string} de busca\footnote{\textit{(“software fairness” OR “fairness in AI” OR “fairness in ML” OR “algorithmic discrimination” OR “algorithmic bias” OR “software discrimination” OR “software bias” OR “fairness in software engineering”) AND (“systematic review” OR “systematic literature review” OR “literature review” OR “survey” OR “mapping study”) AND (“software engineering”)}} foi utilizada nos seguintes repositórios: \textit{Google Scholar}, \textit{ACM Digital Library}, \textit{IEE Xplore}, \textit{Scopus} e \textit{Sage Journals}. As fontes incluídas na busca manual estão presentes na Tabela \ref{tab:fontes_busca_manual}.
\begin{table}[htbp]
    \centering
    \scriptsize
    \begin{tabular}{|l|l|}
        \hline
        \textbf{Categoria} & \textbf{Fonte} \\
        \hline
        Conferência & Foundations of Software Engineering (FSE) \\
        Conferência & Empirical Software Engineering and Measurement (ESEM) \\
        Conferência & FAccT \\
        Conferência & ICSE: Research Papers \\
        Conferência & ICSE: Software Engineering in Society \\
        Conferência & ICSE: CHASE (Cooperative and Human Aspects on Software Engineering) \\
        Conferência & ICSE: FairWare (Software Fairness) \\
        Conferência & ICSE: GE (Gender Equality in Software Engineering) \\
        Conferência & ICSE: NIER (New Ideas and Emerging Results Track) \\
        Conferência & ICSE: RAIE (Responsible AI Engineering) \\
        Conferência & ICSE: SE4RAI (Software Engineering for Responsible AI) \\
        \hline
        Periódico & ACM Transactions on Software Engineering and Methodology (TOSEM) \\
        Periódico & Empirical Software Engineering Journal Issues \\
        Periódico & Information and Software Technology \\
        \hline
    \end{tabular}
    \caption{Lista de fontes da busca manual}
    \label{tab:fontes_busca_manual}
\end{table}
\subsection{Processo de inclusão e exclusão}
Os artigos incluídos (IC1) consistem em revisões de literatura que abordam definições de justiça em engenharia de software, incluindo, mas não se limitando a, fatores que fundamentam essas definições, como essas definições se relacionam com a discriminação algorítmica e os critérios usados para avaliar a justiça. Excluímos estudos que (EC1) não conduzem uma revisão de literatura de justiça de software ou testes de justiça, (EC2) não podem ser acessados através do proxy da Universidade de Calgary, (EC3) não são conduzidos em inglês, (EC4) são duplicatas - mantendo apenas a versão mais recente ou, no caso de um artigo de conferência e sua extensão de periódico, incluindo apenas o último - (EC5) não estão dentro do campo de engenharia de software e (EC6) foram publicados antes de 2017. Ao término da aplicação dos critérios, foram incluídos 60 artigos.

\section{Resultados}
\subsection{Definições de justiça algorítmica}
Muitos estudos têm focado em compreender o conceito de justiça no contexto de Inteligência Artificial (IA) \cite{verma2018fairness, soremekun2022softwarefairnessanalysissurvey, gupta_fairness_2023, rabonato_systematic_2024}. Nesse contexto, a noção de justiça é conceituada como a ausência de viés ou discriminação em decisões tomadas por sistemas digitais \cite{doi:10.3233/EFI-240045}. Justiça algorítmica, ou justiça de software, é classificada como uma propriedade de um software\cite{chen2024fairnesstesting} que busca garantir resultados, ou \textit{outputs} que não discriminem grupos ou indivíduos \cite{doi:10.3233/EFI-240045}. Como algoritmos são construídos em termos matemáticos e justiça é conceituada em função destes algoritmos, justiça também se torna um componente matemático e estatístico \cite{rabonato_systematic_2024, palumbo_objective_2024}, e suas definições e métricas se tornam frequentemente desenvolvidas a partir de uma compreensão de justiça procedural ou baseada na distribuição de bens e resultados.

Justiça procedural consiste no princípio de que todos devem ser tratados iguais durante um processo de decisão \cite{schwobel_long_2022}. Nesse sentido, \textit{fairness through unawareness}, ou justiça através do desconhecimento, considera que um processo de tomada de decisão é justo se não considera atributos sensíveis (como raça e gênero) \cite{schwobel_long_2022, chen2024fairnesstesting}. Em contraponto, \textit{fairness through awareness}, ou justiça através do conhecimento, considera que um sistema é justo caso ele classifique ou tome decisões iguais para indivíduos similares em relação a uma tarefa em particular \cite{dwork2011fairnessawareness}. Dessa maneira, indivíduos considerados similares por um sistema de diagnóstico médico precisam receber o mesmo diagnóstico. A ideia de \textit{fairness through awareness} consiste, então, na concepção de justiça baseada na distribuição de bens e resultados.

Os exemplos analisados acima são fundamentados na compreensão da justiça em nível individual. Em contraponto, justiça em nível de grupo consiste em avaliar as distribuições estatísticas entre diferentes grupos, buscando garantir paridade estatística em tomadas de decisões realizadas em diferentes grupos (homens e mulheres, por exemplo) \cite{dwork2011fairnessawareness}. 

Débito de justiça, ou \textit{fairness debt}, é definido como um conjunto de construções de design, implementação ou gestão que oferecem benefícios de curto prazo, mas estabelecem um contexto técnico que pode tornar mudanças no sistema custosas, resultando em impactos sociais significativos \cite{santos2025softwarefairnessdebt}. Aqui, no entanto, a abordagem sobre justiça algorítmica foca mais em suas causas — como viés histórico — e seus resultados, como danos psicológicos.

\subsection{Métricas de justiça algorítmica}
Diversas formulações são derivadas da concepção de justiça a nível individual e a nível de grupo, constantemente utilizando métricas como verdadeiro positivo, falso positivo, verdadeiro negativo e falso negativo \cite{dwork2011fairnessawareness}. Paridade preditiva, por exemplo, exige que diferentes grupos tenham a mesma taxa de valor preditivo positivo (VPP). Ou seja, que diferentes grupos tenham, entre si, a mesma porcentagem de casos verdadeiros positivos em relação à quantidade total de casos positivos previstos \cite{verma2018fairness}. As definições e métricas de justiça algorítmica são extensas e não se limitam aos exemplos trazidos, mas aqui propomos apenas uma compreensão dos conceitos que fundamentam o campo, visto que outros trabalhos já revisaram extensivamente as múltiplas definições de justiça algorítmica \cite{gupta_fairness_2023, verma2018fairness, dwork2011fairnessawareness, soremekun2022softwarefairnessanalysissurvey, rabonato_systematic_2024, doi:10.3233/EFI-240045}.

A justiça algorítmica, entendida como um atributo do software, leva à criação do teste de justiça. Esse teste envolve qualquer atividade destinada a identificar \textit{bugs} de justiça — imperfeições em um sistema de software que resultam em discrepâncias entre as condições de justiça esperadas e as observadas — por meio da execução do código \cite{chen2024fairnesstesting}. No contexto do aprendizado de máquina, \cite{chen2024fairnesstesting} mapeia e categoriza os testes em cinco tipos: teste de dados, que busca identificar viés nos dados de treinamento; teste do programa de aprendizado de máquina, que avalia falhas no código que podem comprometer a justiça; teste de \textit{frameworks}, voltado para a detecção de falhas estruturais em bibliotecas de aprendizado de máquina; teste do modelo, que verifica o comportamento do modelo com base em suas entradas e saídas; e teste de componentes que não envolvem aprendizado de máquina, que investiga como elementos externos, como armazenamento de dados, podem introduzir vieses no sistema.

\subsection{Procedimentos e técnicas para sistemas de tomada de decisão mais justos}
Tendo definições e métricas de justiça algorítmica estruturadas, os mecanismos para instrumentalizar são classificados em abordagens de pré-processamento, processamento e pós-processamento \cite{fairnessinml_survey}. Abordagens de pré-processamento focam principalmente na instrumentalização de datasets. Datasets desbalanceados são constantemente compreendidos como fatores significativos no processo de discriminação algorítmica, daí torna-se fundamental a construção de datasets compostos por amostras representativas que reflitam de forma equilibrada as características de diferentes grupos sociais, minimizando vieses e promovendo uma distribuição mais justa dos resultados preditivos. Outras abordagens de pré-processamento consistem em ajustar os pesos dos dados (i.e. a importância de cada dado para a determinação do resultado final) \cite{kamiran_data_2012} e realizar transformações probabilísticas nos dados para reduzir a influência de variáveis como gênero e raça nas previsões de modelos \cite{calmon2017optimizeddatapreprocessingdiscrimination}.

Abordagens de processamento focam no algoritmo desenvolvido. \textit{Adversarial learning} \cite{zhang2018mitigatingunwantedbiasesadversarial} é uma técnica em que dois modelos são treinados simultaneamente. O primeiro modelo fica responsável pelas tomadas de decisões e o segundo tenta descobrir características sensíveis, como gênero ou raça. O objetivo é otimizar o modelo responsável pela tomada de decisão e minimizar a capacidade do segundo modelo, adversário, de prever características sensíveis, que podem influenciar em tomadas de decisões discriminatórias. Nessa abordagem, também é possível ajustar algoritmos de tomada de decisão adicionando restrições que se relacionam com as métricas de justiça algorítmica \cite{fairness-aware}. Por exemplo, restrições que garantem que um modelo tenha a mesma taxa de verdadeiros positivos para diferentes grupos, ou que a diferença entre verdadeiros positivos e falsos positivos seja igual entre diferentes grupos \cite{rabonato_systematic_2024}.

Abordagens de pós-processamento consistem em ajustar as decisões resultantes do modelo, que já foi treinado e executado. De maneira similar à técnica de restrição mencionada previamente, \cite{ hardt2016equalityopportunitysupervisedlearning} propõe ajustar previsões geradas através da aplicação algoritmos de otimização, gerando novas previsões em que taxas de falso positivo e falso negativo sejam iguais entre grupos sensíveis. Ferramentas como \textit{FairML} \cite{fairml} contemplam diversas métricas de justiça algorítmica, algoritmos de mitigação de vieses e técnicas para diagnosticar vieses em sistemas de tomada de decisão.

\section{Discussão}
\subsection{Justiça como um atributo de \textit{software}}
A concepção de justiça algorítmica como um atributo de software \cite{soremekun2022softwarefairnessanalysissurvey} impõe alguns limites em relação  às possibilidades de instrumentalização do conceito. Na engenharia de software, justiça é originalmente concebida como uma noção abstrata \cite{schwobel_long_2022}. Isso ocorre porque as dinâmicas de sistemas e algoritmos, ao se manifestarem na sociedade, escapam às representações próprias da engenharia de software. Para que o conceito se adeque aos processos da engenharia de software, este então é adaptado para uma noção formal, matemática e quantificável \cite{rabonato_systematic_2024} de justiça. 

O primeiro erro então é estabelecido: a necessidade de adaptar os impactos sociais de algoritmos aos procedimentos formais da engenharia de software. Tal adaptação, na verdade, distorce a lógica esperada: a engenharia de software, que deveria adaptar-se às dinâmicas sociais que são produzidas através da interação entre seus sistemas e a sociedade, acaba articulando o campo de estudo em questão de maneira isolada, não reconhecendo que esta está em função da sociedade, e não o oposto. O que passa despercebido é que a própria formalização dos impactos sociais de algoritmos e sistemas digitais já é, por si só, uma intervenção da engenharia de software na sociedade. É a própria necessidade de formalizar os impactos sociais de algoritmos que produz a ideia de ``justiça": essa noção já nasce com um caráter limitante porque tem como elemento fundacional as tomadas de decisão realizadas por sistemas algorítmicos. Porém, os impactos sociais de algoritmos não se limitam às suas tomadas de decisão e distribuição de resultados. 

Tomemos como exemplo a Uber: em 2024, motoristas e passageiros durante corridas no Rio de Janeiro, porque o aplicativo sugeriu rotas  que passavam por comunidades dominadas por traficantes \cite{g1_2024, oglobo_2024}. Uma análise dos impactos sociais de algoritmos através das lentes de justiça algorítmica não nos permite investigar os impactos e as dinâmicas sociais das rotas propostas por aplicativos de corrida, quando contextualizadas em relação à segurança pública da cidade.

\subsection{Justiça como um resultado de tomadas de decisões}
O segundo erro consiste então em definir os impactos sociais de algoritmos unicamente a partir de suas tomadas de decisão e distribuição de resultados, pois essa definição falha em reconhecer as características estruturantes de tecnologias na sociedade. Tecnologias de reconhecimento de gênero, como o \textit{genderize}\footnote{\href{http://genderize.io}{\textit{Genderize}} é uma plataforma que utiliza inteligência artificial para supor o gênero atribuído à um nome, de acordo com o país escolhido.}, pressupõem a noção binária de gênero, e simplesmente avaliar quão justas são as predições desses sistemas  ``permite a ignorância das maneiras pelas quais os humanos e a tecnologia co-conspiram para [...] defender e reproduzir ativamente estruturas sociais discriminatórias"\ \cite{Hoffmann07062019}. 

Os impactos e a discriminação algorítmica estão então estritamente relacionados não apenas às suas tomadas de decisões, mas à própria concepção das tecnologias  desenvolvidas, e as abordagens de justiça propostas não apenas falham em mitigar os efeitos das estruturas maiores nas quais os sistemas algorítmicos são implantados \cite{west_redistribution_2020}, como continuam a reproduzi-los. A noção de justiça a partir de resultados de tomadas de decisões significa também que esta se concentra em analisar quais grupos ou indivíduos estão em desvantagem devido às tomadas de decisão, e como \cite{Hoffmann07062019} expõe, a mudança é sutil, mas com consequências: ao focar em uma análise sobre desvantagens, as abordagens de justiça algorítmica não questionam as condições normativas que produzem sujeitos favorecidos.

E mesmo que a definição de justiça algorítmica pudesse ser vista como um produto das decisões tomadas por um sistema, a engenharia de software ainda falharia pela maneira que olha para categorias estruturantes na sociedade, como raça e gênero: pesquisadores constantemente evitam justificar e explicar como adotam diferentes categorias de raça em estudos sobre justiça algorítmica, e que restrições técnicas e os próprios desejos dos pesquisadores levam-os a esquemas de categorização mais simples com menos categorias raciais \cite{abdu_empirical_2023}. Porém:

\begin{quote}
[...] os cientistas da computação não podem escapar do complicado trabalho político de classificação racial ao usar uma definição matemática particular de justiça. \cite{abdu_empirical_2023}
\end{quote}

\subsection{O pensamento computacional e o papel da abstração}
\cite{palumbo_objective_2024} afirma que princípios éticos, devido à sua subjetividade, são de pouca utilidade para cientistas da computação, pois não é possível encontrar maneiras de padronizá-los. A complexidade subjetiva com que algoritmos impactam dinâmicas sociais não é, no entanto, algo a ser combatido e eliminado. Na verdade, esse é o aspecto crucial que deve ser reconhecido caso a computação realmente busque se tornar uma ciência mais consciente dos seus impactos e mais comprometida com questões sociais. Porém, o caminho que parece ser seguido — a necessidade de formalizar matematicamente a justiça — justificada por uma suposta abstração presente no termo \cite{palumbo_objective_2024, schwobel_long_2022}, se contradiz quando justaposta a uma característica chave da computação: o pensamento computacional tem como uma de suas características a abstração. Para simplificar a complexidade de sistemas e tornar a programação mais eficiente, construímos nossas lógicas a partir da abstração de componentes que fazem parte destes próprios sistemas. Esse processo de abstração, porém, não consiste apenas em generalizar regras e construir métodos formais, mas também significa desenvolver novas maneiras de mapear realidades sociais, consequentemente determinando o que deve ser intervindo, e de que maneira \cite{malazita_infrastructures_2019}. 

Não é verdade, então, que a justiça algorítmica nasce como um conceito abstrato, e por isso precisamos formalizá-la. Na verdade, o completo oposto ocorre a partir da seguinte contradição: como já mencionado, a engenharia de software, na tentativa de instrumentalizar os impactos sociais de algoritmos e sistemas digitais, formaliza esses impactos em termos de uma justiça baseada em tomadas de decisões probabilísticas, e é esse exato processo que desconsidera as dinâmicas sociais de algoritmos que surgem quando estes são imersos em estruturas sociais maiores. O maior desafio é compreender que, enquanto buscamos debater os impactos sociais dos algoritmos e aproximar a engenharia de software da sociedade, acabamos perpetuando maneiras de pensar o mundo que, sob a aparência de ``justiça algorítmica”, continuam nos distanciando da sociedade e reforçando uma construção centrada em nossos próprios termos.

\subsection{Possíveis caminhos}
As problemáticas discutidas demonstram alguns dos sintomas de um campo de estudo mais próximo da ética empresarial convencional que de tradições políticas e sociais mais radicais \cite{Greene2019BetterNC}. Não há uma alternativa intuitiva para isso, mas os exemplos da Uber e \textit{genderize} permitem compreendermos que um sistema supostamente justo não necessariamente reflete em impactos e dinâmicas sociais positivas. Talvez seja necessário, então, não apenas questionar se um sistema é justo, mas também de que maneiras ele interage com estruturas sociais maiores e resulta em dinâmicas sociais complexas e prejudiciais. E esse, claro, não é um problema que a computação está apta a resolver por si só: pesquisadores das ciências sociais, comunicação, direito e outras áreas têm constantemente buscado compreender essa questão. \cite{mendonca_algorithmic_2023}, por exemplo, concebe algoritmos como instituições nas sociedades contemporâneas, argumentando que eles não apenas estruturam a tomada de decisões, mas também constroem relações de poder.

Os impactos sociais dos sistemas algorítmicos não podem ser totalmente compreendidos a menos que sejam conceituados dentro de contextos institucionais e sistemas de opressão e controle mais amplos, e isso só é possível ao direcionar o foco não apenas para aqueles impactados negativamente por sistemas digitais e inteligência artificial, mas também para aqueles que detêm poder e se beneficiam dos algoritmos \cite{study_up}. O melhor caminho para a computação e engenharia de software, talvez, não seja a construção de métodos formais de avaliação de justiça, mas sim a condução de um processo no qual o usuário seja também um indivíduo inserido em um contexto social e político. Nesse processo, os requisitos devem ir além dos critérios funcionais e não funcionais, incorporando também compromissos políticos. Os testes de justiça, por sua vez, não devem se limitar à avaliação da validade de uma decisão, mas devem considerar quem realmente se beneficia dela, mesmo quando esta é considerada justa.

\section{Considerações finais}
Nesse trabalho, propomos uma discussão sobre alguns limites das abordagens propostas para a criação e validação de sistemas algorítmicos e de inteligência artificial justas. O campo de estudo sobre justiça algorítmica é recente e, apesar da revisão de literatura realizada indicar uma tendência às abordagens instrumentalistas e formais, não existe consenso na maneira à qual a ideia de justiça algorítmica é operacionalizada. Este torna-se, então, um momento oportuno para investigar os problemas e as contradições emergentes da justiça algorítmica.

Para isso, é fundamental não se ater unicamente ao questionamento sobre como construir algoritmos com tomadas de decisões justas. Compreender que o campo de estudo em questão não existe de forma fixa que precede a própria investigação conduzida por cientistas da computação e engenheiros de software abre possibilidades para outras perguntas: De que maneiras os algoritmos se entrelaçam com estruturas sociais e contribuem para a produção de desigualdades? Quem ganha com a persistência de sistemas algorítmicos injustos e discriminatórios? Como a ciência da computação e a engenharia de software podem trazer esse questionamento às suas práticas? E, sobretudo, o que se ganha e o que se perde ao analisar os impactos dos algoritmos a partir da perspectiva predominante na ciência da computação e engenharia de software?

\clearpage
\bibliographystyle{sbc}
\bibliography{refs}

\end{document}